\documentclass[12pt]{article}
\usepackage{graphicx}
\usepackage{amssymb}
\def\lsim{\mathrel{\raise.3ex\hbox{$<$\kern-.75em\lower1ex\hbox{$\sim$}}}}
\def\gsim{\mathrel{\raise.3ex\hbox{$>$\kern-.75em\lower1ex\hbox{$\sim$}}}}

\begin{document}

\title{Neutral currents and
tests of three-neutrino unitarity in long-baseline experiments}

\author{V. Barger$^{1}$, S. Geer$^{2}$, and K. Whisnant$^{3}$\\[2ex]
\small\it $^1$Department of Physics, University of Wisconsin, Madison, WI 53706, USA\\
\small\it $^2$Fermi National Accelerator Laboratory, P.O. Box 500, Batavia, IL 60510, USA\\
\small\it $^3$Department of Physics, Iowa State University, Ames, IA 50011, USA}

\date{}

\maketitle

\begin{abstract}

We examine a strategy for using neutral current measurements in
long-baseline neutrino oscillation experiments to put limits on the
existence of more than three light, active neutrinos. We determine the
relative contributions of statistics, cross section uncertainties, event
misidentification and other systematic errors to the
overall uncertainty of these measurements.
As specific case studies, we make simulations of beams and detectors that are
like the K2K, T2K, and MINOS experiments.
We find that the neutral current cross section uncertainty and contamination
of the neutral current signal by charge current events allow a sensitivity
for determining the presence of sterile neutinos at the 0.10--0.15 level in probablility.

\end{abstract}

\newpage

\section{Introduction}

In recent years a series of exciting experimental results have shown that neutrinos 
have finite masses and mixings. For a recent review of the status see Ref.~\cite{bmw-review}. Solar neutrino and atmospheric neutrino results  
indicate that all three known neutrino flavors ($e$, $\mu, \tau$) participate in neutrino 
mixing, and hence neutrino oscillations. Consequently, the standard 
framework to describe 
the experimental results and analyse neutrino oscillation data is 
that of three-flavor mixing in which the three flavor eigenstates are related to 
three mass eigenstates by a $3 \times 3$ mixing matrix\cite{mnsp}. The positive signal for
$\bar\nu_\mu\to\bar\nu_e$ oscillations from the LSND 
experiment\cite{lsnd} challenges the three-flavor mixing paradigm\cite{4nu}. However, the neutrino oscillation interpretation of the LSND observations is yet to be confirmed. Independent of whether 
or not the LSND results are confirmed by MiniBooNE\cite{miniboone}, the three-flavor 
mixing framework deserves further experimental scrutiny in the coming years. 
Much of the focus on future experiments so far has been directed to the determination of the 3-mixing angles and the CP-violating phase with long-baseline or oscillation 
experiments\cite{lbl-osc} and reactor experiments\cite{reactor-osc}. Of interest in this paper is the measurement of the neutral current, which could allow tests of the unitarity of the $3\times3$ mixing matrix and thus indirectly probe the existence of sterile neutrinos.

In a three-flavor neutrino model, the sum of the oscillation probabilities\break
$\sum_{y=e,\mu,\tau} P(\nu_x \to \nu_y)$ is unity. If
there are more than three light neutrinos, we know 
from measurements of the invisible width of the $Z$~\cite{pdg} 
that the additional neutrinos must be sterile. 
If additional light neutrinos mix with the 
three known flavors we can expect a non-zero oscillation probability 
to sterile neutrinos, $P(\nu_x \to \nu_s) \ne 0$. To test 
the three-flavor neutrino-mixing paradigm it is important to search for a 
sterile neutrino component within the neutrino flux from natural and 
manmade sources. Since sterile neutrinos have no strong or electroweak
interactions, they cannot be detected directly. However,
neutral current (NC) measurements allow $\sum_{y=e,\mu,\tau} P(\nu_x \to \nu_y)$ 
to be determined which, by 
probability conservation, is equal to $1 - P(\nu_x \to \nu_s)$. 
Therefore, in principle a NC measurement alone 
is sufficient to determine $P(\nu_x \to \nu_s)$. However, in a realistic 
detector misidentifications of 
CC and NC events, together with systematic 
uncertainties on the relevent neutrino interaction cross sections, complicate the 
analysis. 

In this paper we study the use of NC measurements 
to determine limits on the sterile neutrino content in 
long-baseline neutrino oscillation experiments. First we consider the
sensitivity to the sterile content that might be obtained in a K2K-like\cite{k2k}, T2K-like\cite{t2k}, 
and MINOS-like\cite{minos} experiment with a
``perfect'' detector and ``perfect'' beam  if there are no systematic uncertainties. 
We then consider the impact on the sensitivity of 
event misidentification and systematic uncertainties.  
Our study is based on a simple simulation of the long-baseline neutrino beams, 
neutrino interactions~\cite{neugen}, and detector responses. 
We present our results versus event rates and the size of
the cross section uncertainty in order to show the dependence on
these quantities.

\section{Using NC data to determine sterile content}

\subsection{Formalism}

The present and proposed long-baseline neutrino oscillation experiments 
exploit conventional neutrino beams that are produced by the decays of 
charged pions in a long channel. This produces a beam which is 
initially almost entirely $\nu_\mu$. Kaons and muons decaying in the channel 
introduce a small (typically $\sim1$\%) $\nu_e$ component in the neutrino beam. 
As the neutrino beam travels towards a distant detector its flavor content will 
evolve. 
In our analysis we will consider three active neutrinos ($\nu_e$, $\nu_\mu$,
$\nu_\tau$) and one sterile neutrino ($\nu_s$), with oscillation
probabilities $P(\nu_\mu \to \nu_x) \equiv P_{\mu x}$, $x = e, \mu,
\tau, s$. We begin by considering an oscillation experiment that has an initially pure 
$\nu_\mu$ beam with well known neutrino spectrum and flux,
and a detector with perfect identification of the produced events. 
Then the event rates at the far detector will be
\begin{eqnarray}
N_{\rm NC} &=& N_\mu^0 (1-P_{\mu s}) \sigma_{\rm NC} / \sigma_\mu \,,\\
N_\mu &=& N_\mu^0 P_{\mu\mu}\,, \\
N_e &=& N_\mu^0 P_{\mu e} \sigma_e / \sigma_\mu \,, \\
N_\tau &=& N_\mu^0 P_{\mu\tau} \,.
\end{eqnarray}
where $N_\mu^0$ is the predicted number of $\nu_\mu$ CC interactions in the detector 
in the absence of oscillations and the $\sigma_x$ denote the interaction cross sections 
$(\sigma_e, \sigma_\mu, \sigma_\tau)$ for ($\nu_e$, $\nu_\mu$, $\nu_\tau$) CC interactions 
and $\sigma_{\rm NC}$ for $\nu_x$ NC interactions.

In an ideal experiment $N_{\rm NC}$ determines $P_{\mu s}$ and $N_{\rm
CC}$ determines $P_{\mu\mu}$.  In practice the presence of $\nu_e$ and
$\nu_\tau$ CC events complicates the analysis if these events are not
distinguished from NC events. In that circumstance the NC events
provide a measure of $1-P_{\mu s} + \epsilon_e P_{\mu e} +
\epsilon_\tau P_{\mu\tau}$, where the factors $\epsilon_e$ and
$\epsilon_\tau$ reflect the contaminations.

Probability conservation ($P_{\mu e} + P_{\mu\mu} + P_{\mu\tau} = 1$)
can be used to eliminate $P_{\mu\tau}$ or $P_{\mu e}$, but not
both. If the beam energy is below the threshold for $\tau$ production
or the probability $P_{\mu e}$ is small and can be neglected, then
$P_{\mu s}$ can still be determined. However, for a realistic detector
with particle misidentifications and/or a $\nu_e$ component at the far
detector that cannot be neglected, the problem of determining $P_{\mu
s}$ can be complex but still solvable, as we shall discuss.

In general, let the probability that an event of type $x$ (NC, $\nu_e$
CC, $\nu_\mu$ CC, or 
 $\nu_\tau$ CC) be identified in the detector
as an event of type
 $y$ be given by $\zeta_{xy}$, where $x,y = (NC,
e, \mu, \tau)$ (note
 that $\zeta_{xx}$ is the efficiency for
detecting an event of
 type $x$). If $N_\mu^0$ is the predicted
number of $\nu_\mu$ CC
 interactions in the detector in the absence
of oscillations, then
 after including oscillations, detector
efficiencies and
 mis-identifications, and integrating over the
energy dependence, the
 number of measured events of type $y$ will
be:
\begin{equation}
N_y = {N_\mu^0\over\sigma_\mu}
\left[(1-P_{\mu s}) \sigma_{NC} \zeta_{NC,y} + 
\sum_{x=e,\mu,\tau} P_{\mu x} \sigma_x \zeta_{xy} \right] \,,
\label{eq:N}
\end{equation}
where the interaction cross sections for $\nu_e$ CC, $\nu_\mu$ CC, $\nu_\tau$ CC,
and NC events are given by $\sigma_e$, $\sigma_\mu$, $\sigma_\tau$ and
$\sigma_{NC}$, respectively.

\subsection{Ignore $\nu_e$'s}

We consider first the situation in which the $\nu_e$ component in the beam at the far 
detector is so small that $\nu_e$ CC interactions can be neglected. 
In this case we let $P_{\mu e} \to 0$ (it is known to be small, at the 5\%
level or less from the CHOOZ experiment\cite{chooz}). 
It is convenient to define the following two ratios,
\begin{eqnarray}
R_{NC} &\equiv& {N_{NC}\over \zeta_{NCNC}N_{NC}^0} =
(1 - P_{\mu s}) + f_{\mu,NC} P_{\mu\mu} + f_{\tau,NC} P_{\mu\tau} \,,
\label{eq:RNC}
\\
R_\mu &\equiv& {N_\mu\over \zeta_{\mu\mu}N_\mu^0} = 
f_{NC,\mu} (1 - P_{\mu s}) + P_{\mu\mu} + f_{\tau,\mu} P_{\mu\tau} \,,
\label{eq:Rmu}
\end{eqnarray}
where
\begin{equation}
N_x^0 \equiv \sigma_x N_\mu^0/\sigma_\mu\,,
\end{equation}
 and
\begin{equation}
f_{x,y} \equiv \zeta_{xy}\sigma_x/\zeta_{yy}\sigma_y 
\label{eq:f}
\end{equation}
is a normalized misidentification factor that gives the fraction of events identified 
as being of type $y$ that are really events of type $x$.
Measuring $R_{NC}$ and $R_\mu$ is
sufficient for deducing $P_{\mu s}$ (and $P_{\mu\mu}$). The analysis
depends on whether or not we are above the $\nu_\tau$ CC interaction threshold, i.e., whether
or not there are $\nu_\tau$ CC events produced in the detector.

\subsubsection{Below $\tau$ threshold}

For neutrino energies below the $\tau$ threshold 
$\sigma_\tau=0$ and $f_{\tau,j} = 0$. In this case we can invert
Eqs.~\ref{eq:RNC} and \ref{eq:Rmu} to obtain
\begin{eqnarray}
P_{\mu\mu} &=& {R_\mu - R_{NC} f_{NC,\mu}\over
1-f_{\mu,NC}f_{NC,\mu}} \,,
\label{eq:Pmm}
\\
P_{\mu s} &=& 1 - {R_{NC} - R_\mu f_{\mu,NC}\over
1-f_{\mu,NC}f_{NC,\mu}}.
\label{eq:Pms}
\end{eqnarray}
Adding uncertainties in quadrature we get
\begin{eqnarray}
\delta P_{\mu\mu} &=& {\sqrt{ (\delta R_\mu)^2
+ f_{NC,\mu}^2 (\delta R_{NC})^2} \over
1 - f_{\mu,NC}f_{NC,\mu}}\,,
\\
\delta P_{\mu s} &=& {\sqrt{ (\delta R_{NC})^2
+ f_{\mu,NC}^2 (\delta R_\mu)^2} \over
1 - f_{\mu,NC}f_{NC,\mu}}\,,
\end{eqnarray}
where in the limit of Gaussian statistical uncertainties 
\begin{equation}
\delta R_j = R_j \sqrt{ {1\over N_j} + \epsilon_j^2 } \,,
\label{eq:deltaR}
\end{equation}
and
\begin{equation}
\epsilon_j \equiv \delta N^0_j/N^0_j\,.
\end{equation}
The first term in each $\delta R_j$ is the usual statistical
uncertainty, the second comes from the normalization uncertainty (flux
{\it and} cross section).

Note that the normalized mis-identification factors $f_{\mu,NC}$ and 
$f_{NC,\mu}$ will be sensitive to the neutrino energy spectrum and the 
detector technology, and therefore must be evaluated for each experimental setup. 
Most of the mis-identification terms are suppressed by
$f^2$; if $f \le 0.1$ 
then $f^2 \le 0.01$.  If the experimental setup is such that 
we can ignore all terms of order $f^2$, 
(see Table~1) 
we have
\begin{equation}
{P_{\mu s}\over\delta P_{\mu s}} \simeq
{1 - R_{NC} + f_{\mu,NC}R_\mu \over \delta R_{NC}} \,,
\label{eq:fom}
\end{equation}
which measures the significance of the deviation of $P_{\mu s}$ from
zero.

For a perfect detector that can identify each event correctly, $f_{x,y} =
\delta_{xy}$. In this limit $P_{\mu s} = 1 - R_{NC}$ and
\begin{equation}
{P_{\mu s}\over\delta P_{\mu s}} \simeq
{P_{\mu s} \over \sqrt{(1-P_{\mu s}){1\over\zeta_{NCNC}N^0_{NC}}
+ (1-P_{\mu s})^2 \epsilon_{NC}^2} } \,.
\label{eq:perfect}
\end{equation}
This ratio depends only on $P_{\mu s}$, the experimental statistics, and the
systematic uncertainty on the NC measurement. Thus, Eq.~\ref{eq:perfect}
defines the maximum sensitivity that is in principle achievable for a given $N^0_{NC}$ and
$\epsilon_{NC}$.

\subsubsection{Above $\tau$ threshold}

If the neutrino energy is above the $\tau$ threshold and there is not a clean signature for
$\nu_\tau$ CC events, we can still deduce $P_{\mu\mu}$ and
$P_{\mu s}$ by using the identity $P_{\mu\mu} + P_{\mu\tau} + P_{\mu
s} = 1$ to eliminate $P_{\mu\tau}$ in Eqs.~\ref{eq:RNC} and
\ref{eq:Rmu} (we are still assuming $P_{\mu e}=0$), which gives
\begin{eqnarray}
P_{\mu\mu} &=& {R_\mu(1+f_{\tau,NC}) +R_{NC}(f_{NC,\mu}+f_{\tau,\mu})
\over 1 + f_{\tau,NC} - f_{\tau,\mu} + f_{\tau,NC}f_{NC,\mu}
-f_{\mu,NC}(f_{NC,\mu}+f_{\tau,\mu})} \,,
\label{eq:Pmm2}
\\
P_{\mu s} &=& 1 - {R_{NC}(1-f_{\tau,\mu}) - R_\mu(f_{\mu,NC}-f_{\tau,NC})
\over 1 + f_{\tau,NC} - f_{\tau,\mu} + f_{\tau,NC}f_{NC,\mu}
-f_{\mu,NC}(f_{NC,\mu}+f_{\tau,\mu})} \,. 
\label{eq:Pms2}
\end{eqnarray}
If no other process contaminates the $\nu_\mu$ CC events (i.e.,
$f_{j,\mu} = 0$ as appears to be the case for a MINOS-like experiment; see
Sec.~\ref{sec:detsim}), then
\begin{equation}
{P_{\mu s} \over \delta P_{\mu s}} \simeq {1 + f_{\tau,NC} - R_{NC}
+ R_\mu (f_{\mu,NC} - f_{\tau,NC}) \over
\sqrt{(\delta R_{NC})^2 + (f_{\mu,NC} - f_{\tau,NC})^2 (\delta R_\mu)^2} } \,.
\label{eq:fom2}
\end{equation}
For a perfect detector, $P_{\mu s}/\delta P_{\mu s}$ is again given by
Eq.~\ref{eq:perfect}.

\subsection{Do not ignore $\nu_e$'s}

If the $\nu_e$ CC interaction rate in the far detector is not negligible (which could
be the case if $\sin^22\theta_{13}$ is near its upper bound and we
want to push the uncertainty in the measurement of $P_{\mu s}$ down to
the few per cent level), then we need three measurements to be able to
solve for all of the probabilities. The potential measurables are
\begin{eqnarray}
R_{NC} &\equiv& {N_{NC}\over \zeta_{NCNC}N_{NC}^0} =
(1 - P_{\mu s}) + f_{\mu,NC} P_{\mu\mu}
+ f_{e,NC} P_{\mu e} + f_{\tau,NC} P_{\mu\tau} \,,
\label{eq:RNC2}
\\
R_\mu &\equiv& {N_\mu\over \zeta_{\mu\mu}N_\mu^0} = 
f_{NC,\mu} (1 - P_{\mu s}) + P_{\mu\mu}
+ f_{e,\mu} P_{\mu e} + f_{\tau,\mu} P_{\mu\tau} \,,
\label{eq:Rmu2}
\\
R_e &\equiv& {N_e\over \zeta_{ee}N_e^0} = 
f_{NC,e} (1 - P_{\mu s}) + f_{\mu,e} P_{\mu\mu}
+ P_{\mu e} + f_{\tau,e} P_{\mu\tau} \,,
\label{eq:Re2}
\end{eqnarray}
and, if we are above the $\nu_\tau$ CC interaction threshold,
\begin{equation}
R_\tau \equiv {N_\tau\over \zeta_{\tau\tau}N_\tau^0} = 
f_{NC,\tau} (1 - P_{\mu s}) + f_{\mu,\tau} P_{\mu\mu}
+ f_{e,\tau} P_{\mu e} + P_{\mu\tau} \,.
\label{eq:Rtau2}
\end{equation}

\subsubsection{Below $\tau$ threshold}
\label{sec:2.3.1}

Below the $\nu_\tau$ CC threshold energy the three measurements must be $R_\mu$,
$R_{NC}$, and $R_e$. Then $f_{\tau,j}=0$, the
$P_{\mu\tau}$ terms drop out, and we can invert
Eqs.~\ref{eq:RNC2}-\ref{eq:Re2} to obtain
\begin{eqnarray} 
&&\hspace{-4em} P_{\mu\mu} = {R_\mu (1 - f_{e,NC}f_{NC,e})
- R_{NC} (f_{NC,\mu} - f_{NC,e}f_{e,\mu}) 
- R_e (f_{e,\mu} - f_{e,NC}f_{NC,\mu}) \over 1-f} \,,
\label{eq:Pmm3}
\\
&&\hspace{-4em} P_{\mu e} = {R_e (1 - f_{\mu,NC}f_{NC,\mu})
- R_{NC} (f_{NC,e} - f_{NC,\mu}f_{\mu,e}) 
- R_{\mu} (f_{\mu,e} - f_{\mu,NC}f_{NC,e}) \over 1-f} \,,
\label{eq:Pme3}
\\
&&\hspace{-4em} P_{\mu s} = 1 - {R_{NC} (1 - f_{\mu,e}f_{e,\mu})
- R_\mu (f_{\mu,NC} - f_{\mu,e}f_{e,NC}) 
- R_e (f_{e,NC} - f_{e,\mu}f_{\mu,NC}) \over 1-f} \,,
\label{eq:Pms3}
\end{eqnarray}
where $f\equiv f_{\mu,NC}f_{NC,\mu} + f_{e,NC}f_{NC,e} +
f_{\mu,e}f_{e,\mu} - f_{\mu,NC}f_{NC,e}f_{e,\mu} -
f_{NC,\mu}f_{\mu,e}f_{e,NC}$. The calculation of the $\delta P$'s is
straightforward; each $R$ term has a statistical and systematic
uncertainty given by Eq.~\ref{eq:deltaR}.

Note that for an idealized detector in which no other processes significantly contaminate 
$\nu_e$ CC events (i.e., $f_{j,e} \simeq 0$) and $\nu_e$ CC events do not
contaminate $\nu_\mu$ CC events (i.e., $f_{e,\mu} \simeq 0$), 
then $P_{\mu e} = R_e$. Since $P_{\mu e}$ is small (of order 0.1 or
less, as indicated by current oscillation limits), eliminating terms
of order $f^2$ and $f P_{\mu e}$ in this case will recover the situation
where we ignored $\nu_e$ (i.e., Eq.~\ref{eq:fom}).

\subsubsection{Above $\tau$ threshold, no $\tau$ measurement}

For energies above the $\nu_\tau$ CC interaction threshold the $P_{\mu\tau}$ terms do not drop
out of Eqs.~\ref{eq:RNC2}--\ref{eq:Re2}. If we do not have the means to
measure $\nu_\tau$ CC events but can measure $\nu_e$ CC events, then
we can use probability conservation to eliminate $P_{\mu\tau}$, giving
\begin{eqnarray}
R_{NC} &=&
(1 - P_{\mu s})(1+f_{\tau,NC}) + P_{\mu\mu}(f_{\mu,NC}-f_{\tau,NC})
+ P_{\mu e}(f_{e,NC}-f_{\tau,NC}) \,,
\label{eq:RNC3}
\\
R_\mu &=&
(1 - P_{\mu s})(f_{NC,\mu}+f_{\tau,\mu}) + P_{\mu\mu}(1-f_{\tau,\mu})
+ P_{\mu e}(f_{e,\mu} - f_{\tau,\mu}) \,,
\label{eq:Rmu3}
\\
R_e &=&
(1 - P_{\mu s})(f_{NC,e}+f_{\tau,e}) + P_{\mu\mu}(f_{\mu,e}-f_{\tau,e})
+ P_{\mu e}(1-f_{\tau,e}) \,.
\label{eq:Re3}
\end{eqnarray}
The general solution for the probabilities is somewhat messy, but if we
assume that no other processes contaminate the $\nu_\mu$ CC signal
(i.e., $f_{j,\mu} \simeq 0$) and the $\nu_e$ CC events do not contaminate
the other signals ($f_{e,j} \simeq 0$), 
(see Sec.~\ref{sec:detsim}), then $P_{\mu\mu} = R_\mu$ and we
can invert Eqs.~\ref{eq:RNC3} and \ref{eq:Re3} to obtain
\begin{equation}
P_{\mu s} = 1 - {R_{NC}(1-f_{\tau,e}) + R_e f_{\tau,NC}
+ R_\mu \left[ f_{\tau,NC}(1-f_{\mu,e}) - f_{\mu,NC}(1-f_{\tau,e}) \right]
\over 1 + f_{\tau,NC}(1+f_{NC,e}) - f_{\tau,e}} \,.
\label{eq:Pms4}
\end{equation}
The calculation of $\delta P_{\mu s}$ is straightforward.

If no other processes contaminate the $\nu_e$ CC signal (i.e., $f_{j,e}
\to 0$), then $P_{\mu e} = R_e$ and we obtain
\begin{equation}
{P_{\mu s} \over \delta P_{\mu s}} = 
{1 + f_{\tau,NC} - R_{NC} + R_\mu(f_{\mu,NC}-f_{\tau,NC}) - R_e f_{\tau,NC}
\over \sqrt{(\delta R_{NC})^2 + (f_{\mu,NC}-f_{\tau,NC})^2 (\delta R_\mu)^2
+ f_{\tau,NC}^2 (\delta R_e)^2} } \,.
\label{eq:fom3}
\end{equation}
%
%
%
%
\subsubsection{Above $\tau$ threshold with a $\tau$ measurement}

If $R_\tau$ is also measured, in addition to $R_e$, then there are
four measurements ($R_\mu$, $R_{NC}$, $R_e$ and $R_\tau$), but
there are only three independent quantities (since $P_{\mu\mu} +
P_{\mu e} + P_{\mu\tau} + P_{\mu s} = 1$). One possible approach would
be to assume that $P_{\mu s}$ is independent of the other
probabilities and use these four measurements to test probability
conservation. We do not pursue this option here. Instead, we use
probability conservation to eliminate one of the probabilities and use
three of the four measurements to determine $P_{\mu s}$ (the fourth
measurement could be used to check probability conservation
afterwards). Since $P(\nu_\mu \to \nu_\tau)$ is most likely much
larger than $P(\nu_\mu \to \nu_e)$ in the $L/E$ regime we are
considering, we use $R_\tau$ as the third measurement (along with
$R_\mu$ and $R_{NC}$).  Then the appropriate formulas for the
measurables $R_{NC}$, $R_\mu$ and $R_\tau$ can be found by the  interchange
$\tau \leftrightarrow e$ in Eqs.~\ref{eq:RNC3}--\ref{eq:Re3}.

\section{Detector simulations}
\label{sec:detsim}

We wish to explore how well in principle a neutrino three-flavor
unitarity test can be performed with a given muon-neutrino beam as a
function of dataset size, and study which systematic uncertainties are
likely to be important, and their impact.

We consider first a ``perfect'' experiment in which the sensitivity 
of the unitarity test is determined only by the statistical uncertainties, 
calculated using a parameterization of the known beam flux and
spectrum, together with a simulation of neutrino interactions in the
detector. An event simulation is used to determine the relevent
detection efficiencies and misidentification factors. We use the
NEUGEN Monte Carlo code~\cite{neugen} to simulate neutrino interactions in the
detector. Events are classified as $\nu_\mu$ CC, $\nu_e$ CC, or NC. 
In practice the requirements used to identify events of a given type 
will depend upon the detector technology.
For example, for a water cherenkov detector in our simple analysis 
we will define a $\nu_e$ CC event 
candidate as an event with an electron candidate above threshold. 
An electron candidate is either a real electron  
or a $\pi^0$ with an
energy exceeding 1~GeV (in which case the two daughter photons from
the high energy $\pi^0$ produce cherenkov rings that overlap in
the detector and cannot be distinguished from a single
electromagnetically showering particle). A NC event candidate would be 
an event containing a $\pi^0$ candidate but no muon candidate, 
where a $\pi^0$ candidate 
has two $e$-like rings above threshold (which come from a $\pi^0$ 
with energy less than 1~GeV). The definition of CC and NC events can 
of course be varied, and then tuned to give favorable values for the 
signal efficiencies and mis-identification factors. Examples are shown 
in Table~1.

\subsection{K2K-like and T2K-like Experiments}
\label{sec:detsimk2k}

To identify the most important systematic uncertainties it is useful
to compare the sensitivity of our ``perfect experiment" with that of a
realistic experiment. 
We begin with the K2K experiment. 
K2K uses a beam from 
the KEK laboratory in Japan. The neutrinos in the KEK beam have a mean energy 
of 1.3~GeV~\cite{k2k}, and the neutrinos travel 250~km to the 
Super-K water Cerenkov detector. A new experiment T2K is being planned that 
will exploit a more intense neutrino source that is presently under construction at Tokai, Japan. 
T2K will also use the Super-K detector, but with a slightly longer baseline 
(300~km) and narrow-band beam with an axis displaced slightly from pointing 
directly at the far detector (an ``off-axis'' beam). 
The real experimental sensitivities of the
K2K and T2K experiments can only be determined by the experimental collaborations. In the
following we use the NEUGEN Monte Carlo program to simulate neutrino 
interactions together with a simple model for the response of a Super-K-like detector. 
Although this is inadequate to precisely predict the
real K2K and T2K sensitivities, it does enable us to identify the dominant
sources of systematic uncertainties, and hence explore how the
experimental results will depend upon the sizes of these
systematics. We use the following parameterization of a Super-K-like detector
response:
\begin{itemize}

\item[(a)] A threshold of 197~MeV/c for the detection and measurement
of muons~\cite{yamada}, and 
100~MeV/c for electrons and $\pi^0$'s.
These thresholds approximate those used for the atmospheric neutrino analysis 
of Super-K~\cite{9803006,yamada}. 

\item[(b)] Energy resolutions given by~\cite{9803006}
\begin{equation}
{\Delta E_{rms} \over E}  =  0.005  +  {0.025 \over \sqrt(E)} \,,
\label{eq:deltaE1}
\end{equation}
for electrons and $\pi^0$'s and
\begin{equation}
{\Delta p_{rms} \over p}  =  0.03 \,,
\label{eq:deltaE2}
\end{equation}
for charged pions and muons.

\end{itemize}
In addition, we use a parametrization of the spectra for the K2K and T2K neutrino beams. 

In our analysis we will use only simulated events with visible
energy greater than 0.1~GeV. For our ``basic'' signals we define a
$\nu_\mu$ CC event candidate as an event with a single muon-like ring, 
a $\nu_e$ CC event candidate as an event with a 
single $e$-like ring, and a NC event candidate as an event with two $e$-like rings, 
which are
assumed to be two photons from a single $\pi^0$ decay. 
Given these definitions, 
the detector efficiencies and mis-identification factors determined from our
simulations are listed in Table~\ref{tab:signals}. As shown
in the table, the efficiencies $\zeta_{jj}$ are of order one-half, and there is no
significant contamination of one signal by another due to
mis-identification. Also shown are the results of a more aggressive
signal definition, where a simulated event with an odd number of $e$-like rings
is labeled as a $\nu_e$ CC event candidate, and the remaining events (those with an
even number of $e$-like rings) are labeled as $\nu_\mu$ CC event candidates 
if they have one or
more $\mu$-like rings or NC if they do not. In this more aggressive
scenario no events are discarded, i.e., all events were used for one
of the targeted signals. Although some of the misidentification
factors are slightly larger for the aggressive scenario, overall they
are not greatly changed, while there is a significant improvement in
the efficiencies for the CC events.

\begin{table}
\vbox{\footnotesize
\tabcolsep=.5em
\caption{\label{tab:signals} Signal efficiencies ($\zeta_{jj}$) and
normalized mis-identification factors ($f_{i,j}$) in selected
long-baseline experiments.}
\vskip0.1in
\begin{tabular}{|l|c|l|l|l|l|l|l|}
\hline\hline
Experiment & $j$ (channel) & Signal & $\zeta_{jj}$ & $f_{NC,j}$ & $f_{\mu,j}$ & $f_{e,j}$ & $f_{\tau,j}$\\
\hline
K2K-like
& NC& two $e$-like, no $\mu$-like      & 0.391 & $-$   & 0.068 & 0.052 & $-$\\
(basic)
&$\mu$& one $\mu$-like, no $e$-like    & 0.520 & 0.087 & $-$   & 0.0007 & $-$\\
& $e$ & one $e$-like, no $\mu$-like    & 0.497 & 0.003 & 0.0004 & $-$   & $-$\\
\hline
K2K-like
& NC&even $e$-like, no $\mu$-like      & 0.437 & $-$   & 0.078 & 0.060 & $-$\\
(aggressive)
&$\mu$&even $e$-like, $\ge1$ $\mu$-like& 0.989 & 0.086 & $-$   & 0.003 & $-$\\
& $e$ & odd $e$-like                   & 0.993 & 0.002 & 0.005 & $-$   & $-$\\
\hline
K2K-like
& NC&even $e$-like, no $\mu$-like      & 0.494 & $-$   & 0.081 & 0.011 & $-$\\
(Gaussian
&$\mu$&even $e$-like, $\ge1$ $\mu$-like& 0.994 & 0.073 & $-$   & 0.0007 & $-$\\
beam)
& $e$ & odd $e$-like                   & 0.999 & 0.0003 & 0.0004 & $-$   & $-$\\
\hline
T2K-like
& NC&even $e$-like, no $\mu$-like      & 0.420 & $-$   & 0.25  & 0.006 & $-$\\
(Gaussian
&$\mu$&even $e$-like, $\ge1$ $\mu$-like& 0.988 & 0.036 & $-$   & 0.0014 & $-$\\
beam)
& $e$ & odd $e$-like                   & 0.944 & 0.00002 & 0.00001 & $-$   & $-$\\
\hline
MINOS-like
& NC & no $\mu >$~1~GeV           & 1.000 & $-$   & 0.903 & $-$   & 0.429\\
$P_{\mu e}=0$
& $\mu$ & any $\mu >$~1~GeV       & 0.749 & 0     & $-$   & $-$   & 0\\
\hline
MINOS-like
& NC & no $e$, $\mu$, or $\gamma$ & 0.520 & $-$   & 1.067 & 0.005 & 0.347\\
$P_{\mu e}\ne0$
& $\mu$ & any $\mu >$~1~GeV       & 0.749 & 0     & $-$   & 0     & 0\\
& $e$   & no $\mu >$~1~GeV, $\ge1$ $e$ or $\gamma$ & 0.999 & 0.125 & 0.090 & $-$   & 0.064\\
\hline
\end{tabular} }
\end{table}

To investigate whether our analysis is 
sensitive to the assumed details of the neutrino spectrum  
we have repeated the calculation of efficiencies and misidentification
factors for a K2K-like experiment with a beam that has the same average
energy and beam spread as the K2K beam, but with a Gaussian energy spectrum (no long 
high-energy tail). For the Gaussian beam, the
misidentification factors involving $\nu_e$ were greatly reduced
(since backgrounds from the high-energy tail are now suppressed), but
$f_{\mu,NC}$ and $f_{NC,\mu}$ were only slightly affected.  Since
$f_{\mu,NC}$ is the dominant $f$ factor for a K2K-like experiment, we conclude that
our results are not very sensitive to the detailed beam spectrum we assume.  

We now consider a T2K-like experiment, where we have used a beam spectrum that corresponds to 
a detector 2 degrees off-axis. 
The resulting mis-identification factors for a T2K-like experiment 
are shown in Table~\ref{tab:signals}. 
All of the misidentification factors are reduced except for $f_{\mu,NC}$, which is now 0.25. Therefore, in both the K2K-like and T2K-like experiments, the most important contamination is $\nu_\mu$ CC events being mis-identified as NC events.

\subsection{A MINOS-like experiment}
\label{sec:detsimminos}

The MINOS experiment is a long-baseline oscillation exeriment that will use a neutrino 
beam from the Fermilab Main Injector and an iron-scintillator sampling calorimeter 
730~km away in Minnesota. MINOS is expected to begin data taking early in 2005. 
With a beam energy that 
is about a factor of three higher than the KEK beam, and a 
detector that is very different from the water cerenkov detector used by K2K and T2K,  
the efficiencies and misidentification factors for MINOS will be very different than 
those for the experiments in Japan. To compute the numbers given in Table~\ref{tab:signals} 
we have once again used a parametrization of the neutrino beam spectrum, the NEUGEN Monte 
Carlo Program to simulate neutrino interactions in an iron detector, and a simple 
parametrization of the response of a MINOS-like detector. In particular we assume:

\begin{itemize}

\item[(a)] An energy threshold of 50 MeV for the detection and measurement
of electrons, and charged and neutral pions, and a threshold of 1~GeV for 
the identification and measurement of muons. Note that the MINOS detector is 
expected to be able to determine the charge and measure the momenta of muons 
from 0.5~GeV/c to 100~GeV/c, and to distinguish $\nu_\mu$ CC events from NC events 
if the muons have momenta exceeding about 1~GeV/c~\cite{doug}.

\item[(b)] Energy resolutions given by
\begin{equation}
{\Delta E_{rms} \over E}  =  {0.23 \over \sqrt(E)} \,,
\label{eq:deltaE3}
\end{equation}
for  electrons and $\pi^0$'s,
\begin{equation}
{\Delta E_{rms} \over E}  =  {0.55 \over \sqrt(E)} \,,
\label{eq:deltaE4}
\end{equation}
for charged pions, and 
\begin{equation}
{\Delta p_{rms} \over p}  =  0.05 \,,
\label{eq:deltaE5}
\end{equation}
for muons.
Note that in practice the muon energy resolution for the MINOS experiment is expected to 
be somewhat better (worse) than described by Eq.~\ref{eq:deltaE5} if the muon ranges out 
(does not range out) in the detector. We found that $\Delta p_{rms}/p$ values
as high as 0.10 do not appreciably change our results.
\end{itemize}

As shown in the table, for a MINOS-like experiment there is a very
large contamination of the NC channel by $\nu_\mu$ CC events, and
mis-identification of $\nu_\tau$ CC events as NC events is also
significant. The efficiency for identifying NC events is about
one-half, similar to the K2K-like and T2K-like
experiments\footnote{Although to first order for MINOS all events with
an electron or photon candidate will be classified as NC events, there
are three independent probabilities, and it is necessary to extract a
separate $\nu_e$ signal, in addition to $\nu_\mu$ and NC signals, to
be able to solve for all of the probabilities.  Hence we must try to
select genuine $\nu_e$ interactions from the large NC background. In the
table we also show mis-identification factors when all non-$\mu$
events are classified as NC events, which could be used when $P_{\mu e}$ is very small.}.

\section{Results}

\subsection{A perfect detector}

We first find the sensitivity of the NC unitarity test for a perfect
detector, i.e., a detector that can categorize each event correctly as
CC muon or NC, with no mis-identification and 100\% efficiency. The
figure of merit for a perfect detector is given by
Eq.~\ref{eq:perfect} with $\zeta_{NCNC}=1$.
We show the $3\sigma$ sensitivity for $P_{\mu
s}$ (i.e., the minimum value of $P_{\mu s}$ for which $P_{\mu s} =
3\delta P_{\mu s}$) versus $N_\mu^0$ (the number of CC muons expected
in the detector with no oscillations) for several values of the NC
systematic error in Fig.~\ref{fig:discovery-k2k} (the dotted curves).
At low statistics the sensitivity is very poor, and for high
statistics the sensitivity approaches the asymptotic limit of $3
\epsilon_{NC}/(1 + 3\epsilon_{NC})$, where 
$\epsilon_{NC} \equiv \delta N_{NC}^0 / N_{NC}^0$ is the fractional NC 
normalization uncertainty.

\subsection{More realistic K2K-like and T2K-like experiments}

\begin{figure}[ht]
\centering\leavevmode
\includegraphics[width=5in]{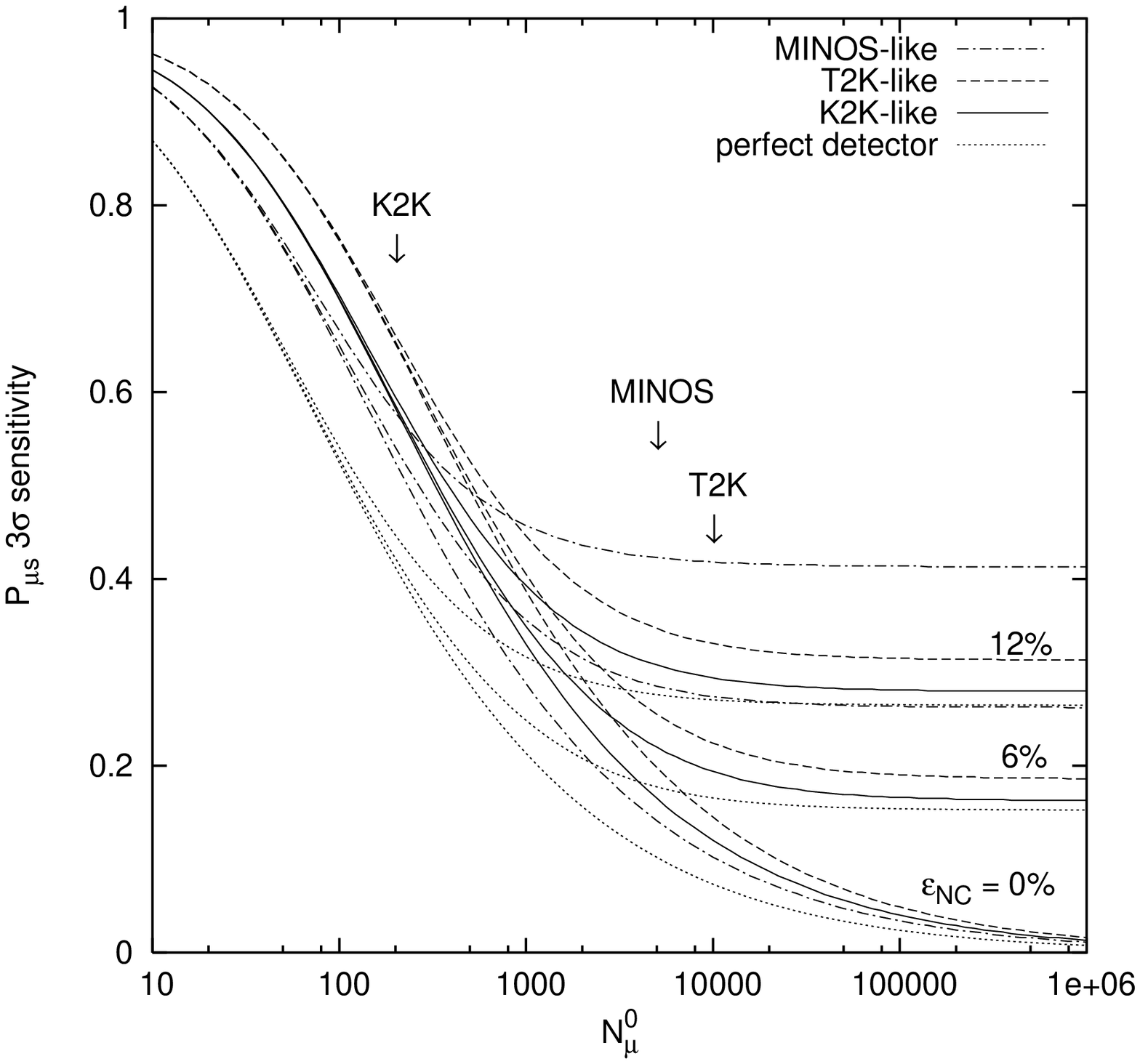}
\caption[]{Assuming Gaussian statistical uncertainties, 
the $3\sigma$ sensitivity for measuring $P_{\mu s}$ versus
$N^0_\mu$ for fixed values of the NC systematic error for a perfect
detector (dotted curves, using Eq.~\ref{eq:perfect}), the K2K-like 
experiment with our basic signal definition (solid), the
T2K-like experiment (dashed), and the MINOS-like experiment (dash-dotted). 
The number of NC events without
oscillations is $N^0_{NC} = 0.156 N^0_\mu$. The systematic uncertainties
$\delta N^0_\mu/N^0_\mu$ and $\delta N^0_e/N^0_e$ are assumed to be
2\%, except when $\epsilon_{NC} \equiv \delta N^0_{NC}/N^0_{NC} = 0$,
in which case they are 0. The arrows indicate the approximate statistical
sensitivities expected for the K2K, T2K and MINOS experiments. 
\label{fig:discovery-k2k}}
\end{figure}

Next we find the NC sensitivity for the K2K-like detector described in
Sec.~\ref{sec:detsimk2k} for the case $P_{\mu e} \simeq 0$. We
generated 400,000 neutrino events using the NEUGEN simulator, from
which the normalized mis-identification factors $f_{x,y}$ were
calculated. For a given set of probabilities $P_{xy}$, the values of
$R_{NC}$ and $R_\mu$ were calculated, and the corresponding
measured value of $P_{\mu s}$ was determined from
Eq.~\ref{eq:Pms}. The uncertainty on $P_{\mu s}$ was calculated using
Eq.~\ref{eq:deltaR}, assuming the uncertainties $\delta R_{NC}$ and
$\delta R_\mu$ are uncorrelated and add in quadrature. The $3\sigma$
sensitivity for $P_{\mu s}$ is shown in Fig.~\ref{fig:discovery-k2k}
(solid curves) for various values of $\epsilon_{NC}$ for the case
$P_{\mu s} = 1 - P_{\mu\mu}$ (all $\nu_\mu$ oscillating to $\nu_s$; we
will consider cases with nonzero $P_{\mu\tau}$ later).  For both low
and high statistics the K2K-like $3\sigma$ sensitivity can be approximated
by
\begin{equation}
P_{\mu s}^{min} \simeq {3 (1 + f_{\mu,NC})(\delta R_{NC}/R_{NC}) \over
1 + 3 (1 + f_{\mu,NC})(\delta R_{NC}/R_{NC})} \,,
\label{eq:3sigk2k}
\end{equation}
which can be derived from Eq.~\ref{eq:fom}, 
where factors quadratic in the $f_{x,y}$ are ignored. Since
$f_{\mu,NC} \simeq 0.08$ for our K2K-like experiment, 
the NC sensitivity is at most about
1.08 worse than that of the perfect detector
for large numbers of events where the statistical uncertainty becomes negligible compared to the systematic uncertainty. At low statistics the efficiency becomes important and the K2K-like performance will be more than 1.08 worse than a perfect detector. 

The K2K-like curves in Fig.~\ref{fig:discovery-k2k} are plotted for the
simple K2K signals in Table~\ref{tab:signals}. The corresponding
curves for the more aggressive K2K-like signals are very similar to the simple
case; the improved efficiencies are partially compensated for by the
slightly higher value of $f_{\mu,NC}$. Thus the result is fairly
insensitive to the exact signal criteria used.

We next consider the effects of nonzero $P_{\mu\tau}$. If we assume
$P_{\mu\tau} = 1 - P_{\mu s}$ (i.e., $P_{\mu\mu} = P_{\mu e} = 0$),
the curves are very close to those of the perfect detector, since the
dominant mis-identification term $f_{\mu,NC}$ does not contribute to
$R_{NC}$ when $P_{\mu\mu} = 0$. If both $P_{\mu\mu}$ and $P_{\mu\tau}$
are both nonzero (with $P_{\mu e} \simeq 0$), the results will lie
somewhere between the curves for K2K-like and the perfect detector.

Finally, we consider nonzero $P_{\mu e}$, in which case $R_e$ must
also be measured and $P_{\mu s}$ is determined using
Eq.~\ref{eq:Pms3}.  As discussed in Sec.~\ref{sec:2.3.1}, if $P_{\mu
e}$ is of order 0.1 or less (as indicated by oscillation bounds such
as from the CHOOZ reactor), and if the misidentification factors are
also of order 0.1 or less, then this case reduces to that where the
$\nu_e$ are ignored. We have verified this numerically for the K2K-like 
misidentification factors in Table~\ref{tab:signals}.

In summary, the sensitivity of the K2K-like detector to the NC signal is
only slightly worse than that of a perfect detector, with the dominant
loss of sensitivity coming from the mis-identification of CC muon
events as NC.
For comparison, in Fig.~\ref{fig:discovery-k2k} we have also shown
sensitivity curves for the T2K-like experiment with a Gaussian beam
spread. Since $f_{\mu,NC} = 0.25$ in this case, the sensitivity is
about a factor of 1.25/1.08 = 1.16 worse than for K2K.

\subsection{The MINOS-like detector}

For the MINOS-like case, we generated 320,000 neutrino events using the NEUGEN simulator, and calculated the corresponding mis-identification factors. The $3\sigma$ sensitivity for $P_{\mu s}$ was calculated as described above for the case $P_{\mu s} = 1 - P_{\mu\mu}$; the results are shown in Fig.~\ref{fig:discovery-k2k}. At low statistics, the MINOS-like experiment does better than the K2K-like and T2K-like experiments because of the higher NC efficiency, but at high statistics it does worse because of the larger mis-identification factors.

\subsection{Exclusion limit when $P_{\mu s} = 0$}

\begin{figure}[ht]
\centering\leavevmode
\includegraphics[width=5in]{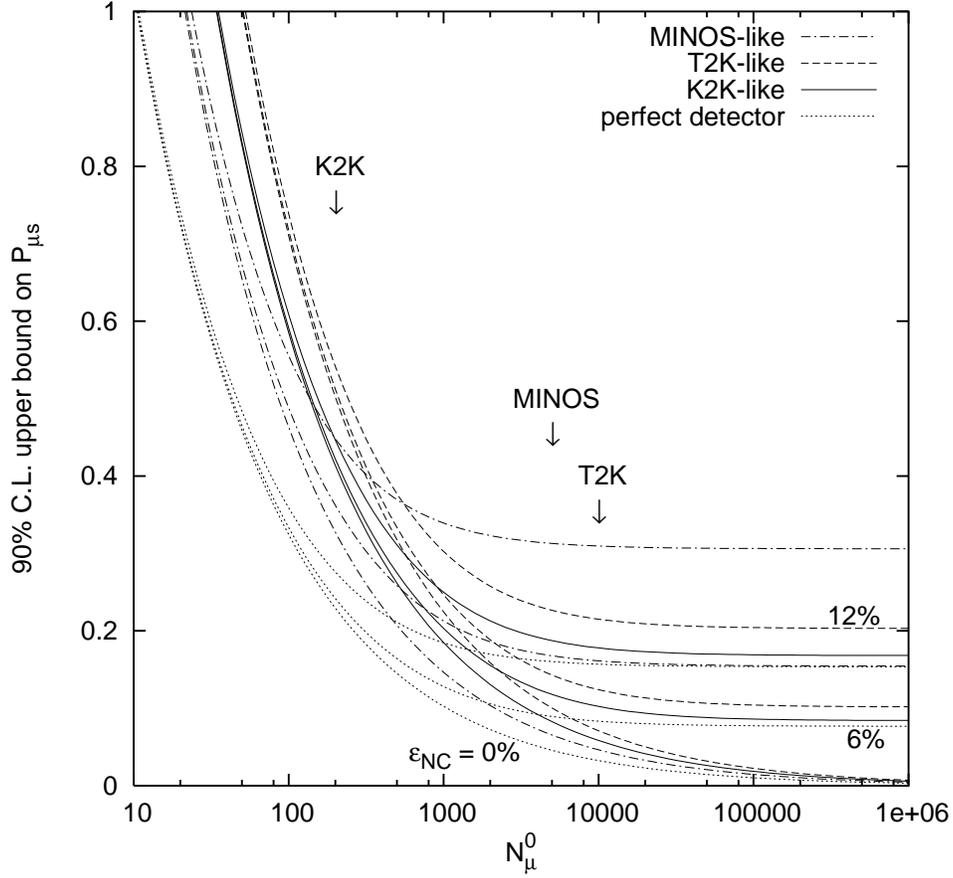}
\caption[]{Assuming Gaussian statistical uncertainties, the 
90\%~C.L. exclusion limit for $P_{\mu s}$ versus $N^0_\mu$
for fixed values of the NC systematic error for a perfect detector (dotted
curves), the K2K-like experiment with basic signals (solid), the T2K-like
experiment with basic signals (dashed), and the MINO-like experiment (dash-dotted).
Other assumptions are the same as
in Fig.~\ref{fig:discovery-k2k}.
\label{fig:exclusion-k2k}}
\end{figure}

If a $3\sigma$ signal for $P_{\mu s}$ is not observed, then an
exclusion limit (upper bound) for $P_{\mu s}$ can then be obtained.
The 90\%~C.L. exclusion limit for $P_{\mu s}$ is shown in
Fig.~\ref{fig:exclusion-k2k} for a perfect detector (dotted curves),
K2K-like with basic signals (solid curves), and T2K-like with basic signals
(dashed curves). To model realistic oscillation probabilities we have
assumed a three-neutrino model assuming the parameters $\delta
m^2_{31} = 2.0\times10^{-3}$~eV$^2$, $\sin^22\theta_{23} = 1.0$, and
$\sin^22\theta_{13} = 0.1$. At high statistics the relative values of the exclusion limits
are approximately proportional to $(1 + f_{\mu,NC})$, similar to the
$3\sigma$ sensitivity levels calculated previously. As was the case for the $3\sigma$ sensitivity, the MINOS-like detector does better than the K2K-like and T2K-like detectors at low statistics, due to the higher NC efficiency, but not as well at high statistics due to larger misidentification factors.

\section{Summary}

At low statistics ($\lesssim 1000$~events), experiments with a larger NC efficiency, such as
our MINOS-like example, tend to have better sensitivity to the sterile
oscillation probability $P_{\mu s}$. At high statistics, the
sensitivity in the cases we considered is primarily limited by the
systematic uncertainty in the NC rate, $\epsilon_{NC}$, and the
contamination of the NC signal from CC $\mu$ events, $f_{\mu,NC}$ (and
NC contamination from CC $\tau$ events, $f_{\tau,NC}$, above $\tau$
threshold). The best anticipated $\epsilon_{NC}$ is of order a few per
cent, so the best $3\sigma$ sensitivity and 90\%~C.L. exclusion limits
that can be expected for the sterile oscillation probability  will be of order 
0.10--0.15 (0.2--0.3 for the oscillation
amplitude). The lowest contamination rates are realized for the K2K-like
and T2K-like cases. Significant improvements in these sterile probability sensitivies or limits can only be achieved by lowering the uncertainty in NC cross sections  or improving the
event selection criteria, both of which could prove to be challenging but very worthwhile.

\section*{Acknowledgments}

We thank D. Harris for a critical reading of the manuscript and H. Gallagher for assistance with NEUGEN.
VB thanks the Aspen Center for Physics for hospitality during the completion of this paper.
This research was supported in part by the U.S. Department of Energy
under Grant Nos.~DE-FG02-95ER40896, DE-AC02-76CH03000 and
DE-FG02-01ER41155, and in part by the Wisconsin Alumni Research
Foundation. 


\newpage

\newpage

\end{document}